\documentclass[12pt]{iopart}
\usepackage{setstack}
\usepackage{amsthm}
\usepackage{iopams}

\newtheorem{thm*}{Theorem}[section]
\newtheorem{defn*}{Definition}[section]
\newcommand{\Ad}{\mathrm{Ad}}
\newcommand{\ad}{\mathrm{ad}}

\begin{document}

\title[On detection of quasiclassical states]{On detection of
quasiclassical states}
\author{Micha{\l{}} Oszmaniec$^1$ and Marek Ku\'{s}$^2$}

\address{Center for Theoretical Physics, Polish Academy of Sciences, Al.
Lotnik\'ow 32/46, 02-668 Warszawa, Poland}

\eads{$^1$ \mailto{michal.oszmaniec@gmail.com}, $^2$ \mailto{marek.kus@cft.edu.pl}}

\begin{abstract}
We give a criterion of classicality for mixed states in terms of expectation
values of a quantum observable. Using group representation theory we identify
all cases when the criterion can be computed exactly in terms of the spectrum
of a single operator.


\end{abstract}
\submitto{\JPA}

\section{Introduction}

Group theoretical coherent states are generalizations of the well known
coherent states of a quantum harmonic oscillator \cite{Perelomov}. Loosely
speaking they appear when an underlying symmetry group $K$ (which is usually
a Lie group) is represented in some Hilbert space $V$. The action of $K$ on a
vector induces an overcomplete set of vectors in $V$. Standard coherent
states of a harmonic oscillator are obtained from the action of the
Heisenberg group in $L^{2}(\mathbb{R})$. The set of coherent states possess
also many other classical features \cite{Perelomov,Classical quantum states}
provided the group $K$ is semisimple, compact and one considers coherent
states that contain the vector of the highest weight. We shall call such
states quasiclassical or, for shortness, simply `classical'. They find
applications in many branches of modern quantum physics ranging from quantum
optics, quantum statistical mechanics, quantum chaos to investigations of
classical limit of quantum systems \cite{CoherentStates}.

The `classicality' of coherent states can manifest itself in several
different aspects. Probably the most important facet is that they minimize
appropriate uncertainty relations \cite{Perelomov,Delbourgo} and as such they
resemble in the most optimal way classical distributions in the phase space.
In the theory of entanglement separable states of multicomponent systems,
i.e., states exhibiting only classical correlations can be also treated as
coherent states for groups of `local' transformations. Adapting the notion of
`locality' to a situation at hand, which on the formal level reduces to a
proper choice of an underlying group and its representation, it is possible
to give an unified treatment of entanglement not only for distinguishable
particles but also for bosonic and fermionic systems \cite{Classical quantum
states,Kotowscy}.

In the general case of semisimple compact groups it is fairly easy to
characterize the set of pure classical \textit{via} a simple algebraic
criterion \cite{Classical quantum states,Lichtenstein}. It can be cast in the
form of the expectation value of an observable (a Hermitean operator) and as
such can be, in principle, a basis for experimental check of classicality.

The concept of classical states can be extended onto mixed states by taking
convex combinations of pure classical states \cite{Classicality of spin
states}. For mixed states the problem of finding sufficient and necessary
criteria of classicality remains essentially open. In fact for separable
states as well as for spin coherent states such criteria are available only
in low-dimensional cases \cite{Classicality of spin states,wootters,Bosons
and Fermions}. The common origin of these criteria, as well as reformulation
of them in terms of expectation values of observables was discussed in
\cite{Classical quantum states} and explained in terms of properties of
generalized coherent states.

The aim of the paper is to provide methods and framework to quantify the
classicality of mixed states from the perspective of the representation
theory of semisimple Lie groups and to give a group theoretic
characterization of cases when it is possible to give an explicit, closed
form criterion for a mixed state to be classical.

The article is organized as follows. In the following section we introduce
some basic notions of representation theory of semisimple Lie groups. In the
third section we define classical states and present their purely algebraic
characterization due to Lichtenstein \cite{Lichtenstein}. Subsequently in
Section 4 we recall briefly Jamio\l{}kowski-Choi isomorphism which connects
completely positive maps on $V$ and nonnegative operators on $V\otimes V$. We
shall make use of this formalism in the rest of the paper. In Section 5 we
define classical mixed states and extend an algebraic criterion of
classicality on this class of states with the help of the {}``convex roof''
construction. We also discuss known examples in which it is possible to
compute appropriate convex roof exactly. In each of those examples exact
computation is possible because classicality of pure states is described by
an equation involving particular antiunitary operator. In Section 6 we
generalize these cases by introducing {}``antiunitary operators detecting
classicality''. This part contains our main findings, we characterize cases
in which such antiunitaries can occur. Firstly, they appear if and only if a
symmetric product $V\vee V$ decomposes onto two irreducible representations
from which one is a trivial, one-dimensional one. Secondly, it turns out that
there is a relation between the existence of such antiunitaries and the
existence of an epimorphism (i.e. an onto homomorphism) of the group $K$ onto one of three groups:
special orthogonal group $SO(N)$ (here $N=\dim(V)$), exceptional Lie group  $G_2$ \cite{Adams} or  $Spin(7)$.

\section{Elements of representation theory}
Let us start with collecting some basic notions and facts from the
representation theory of Lie groups and Lie algebras (see eg.,
\cite{Hall,Brut Raczka}).

In the following $K$ is a compact, simply connected and semisimple Lie group,
and $\mathfrak{k}$ its (real) Lie algebra (compact and semisimple). Due to
simple connectedness of $K$ representations of $K$ and $\mathfrak{k}$ are in
one to one correspondence. Due to compactness of $K$ every complex
representation of $K$ (and thus also of $\mathfrak{k}$) is simply reducible.

We denote by $G=K^{\mathbb{C}}$ and $\mathfrak{g=}\mathfrak{k}^{\mathbb{C}}$
the complexifications of the above algebraic structures, thus $\mathfrak{g}$
is a semisimple complex Lie algebra. Two properties of representations
mentioned above are also inherited by $G$ and $\mathfrak{g}$.

Let $\mathfrak{t}\subset\mathfrak{k}$ be some (fixed) Cartan subalgebra of
$\mathfrak{k}$ i.e., a maximal abelian subalgebra of $\mathfrak{k}$ having
self-normalizing property, (for $T\in\mathfrak{t}$ we have that $[T,\,
X]\in\mathfrak{t}$ implies $X\in\mathfrak{t}$). Let also
$\mathfrak{h}=\mathfrak{t}^{\mathbb{C}}$ be the complexification of
$\mathfrak{t}$.

The group $G$ acts in a natural way on its algebra \textit{via} the adjoint
representation $\Ad:G\rightarrow GL(\mathfrak{g})$, $\Ad_{g}(X)=gXg^{-1}$,
$g\in G$, $X\in\mathfrak{g}$. By differentiation it induces the adjoint
representation of $\mathfrak{g}$,
$\ad:\mathfrak{g}\rightarrow\mathfrak{gl}(\mathfrak{g})$, $\ad_{Y}(X)=[Y,\,
X]$, $X,\, Y\in\mathfrak{g}$. Due to semisimplicity of $\mathfrak{g}$ it
decomposes as
$\mathfrak{g}=\mathfrak{h}\oplus\bigoplus_{\alpha}\mathfrak{g}_{\alpha}$,
where the subspace $\mathfrak{g}_{\alpha}$ is spanned by $X$ such that there
exist an element $\alpha\in\mathfrak{h}^{*}$ (the space dual to
$\mathfrak{h}$) for which $\ad_{H}(X)=\alpha(H)X$. Such
$\alpha\in\mathfrak{h}^{*}$ are called roots whereas $\mathfrak{g}_{\alpha}$
are called root spaces. It turns out that the root spaces are one
dimensional.

It is possible to chose the basis of $\mathfrak{h}^{*}$ (basis elements of
this kind are called positive simple roots) in such a way that all roots can
be expressed either as a positive integer combination of basis elements (such
roots are called positive - $\mathfrak{n}_{+}$ ) or as a negative integer
combination of them (such roots are called negative - $\mathfrak{n}_{-}$).
Therefore the root decomposition of $\mathfrak{g}$ can be rewritten as
$\mathfrak{g}=\mathfrak{n}_{-}\oplus\mathfrak{h}\oplus\mathfrak{n}_{+}$. One
checks that the Lie subalgebras $\mathfrak{n}_{+}$ and $\mathfrak{n}_{-}$ are
nilpotent.

In what follows we denote by $V$ a finite-dimensional complex vector space on
which $K$ and $\mathfrak{k}$ are irreducibly represented \textit{via} complex
representations $\Pi$ and $\pi$, respectively. $\Pi$ and $\pi$ extend
uniquely to representations of $G$ and $\mathfrak{g}$ on $V$ (which will be
denoted by the same symbols).

A convenient way of description of representations of $\mathfrak{g}$ uses the
notion of weights vectors, i.e., simultaneous eigenvectors of representatives
of all elements form the Cartan subalgebra $\mathfrak{h}$. It means that
$v_{\lambda}\in V$ is a weight vector if
$\pi(H)v_{\lambda}=\lambda(H)v_{\lambda}$ for $H\in\mathcal{\mathfrak{h}}$,
where a form $\lambda\in\mathfrak{h}^{*}$ is called the weight of $\pi$.
Since we assume that $\mathfrak{g}$ is semisimple and $V$ is a carrier space
of an irreducible representation, $V$ decomposes as
$V=\oplus_{\lambda}V_{\lambda}$, where summation is over all weights of the
considered representation. The subspaces $V_{\lambda}$ are called weight
spaces and are spanned by the corresponding weight vectors $v_{\lambda}$. An
irreducible representation is uniquely characterized by its highest weight
$\lambda_{0}$ determined by the highest weight vector $v_{\lambda_{0}}$,
i.e., by the (unique, up to multiplicative constant) weight vector
annihilated by all representatives of $n_{+}$,
$\pi(\mathfrak{n}_{+})v_{\lambda_{0}}=0$. We will write $V^{\lambda_{0}}$
instead of $V$ when we want to distinguish which irreducible representation
of $\mathfrak{g}$ is considered.

Since $\mathfrak{g}$ is semisimple its Killing form
$B:\mathfrak{g}\times\mathfrak{g}\rightarrow\mathbb{C}$,
$B(X,Y)=\tr(\ad_{X}\circ\ad_{Y})$ is nondegenerate and establishes a
correspondence between $\mathfrak{g}$ and $\mathfrak{g}^{*}$,
$\tilde{B}:\mathfrak{g}\rightarrow\mathfrak{g}^{*}$,
$\tilde{B}(X)=K(X,\cdot)\in\mathfrak{g}^{*}$, intertwining adjoint and
coadjoint representations, $\Ad_{g}^{*}(\tilde{B}(X))=\tilde{B}(\Ad_{g}(X))$.

Let $X_{i}$ form a basis in $\mathfrak{g}$. An element of the universal
enveloping algebra $\mathfrak{U}(\mathfrak{g})$ (see \cite{Brut Raczka} for
the definition of $\mathfrak{U}(\mathfrak{g})$) defined as
$C_{2}=\sum_{i}\sum_{j}B^{ij}X_{i}X_{j}$, is called the second-order Casimir
operator of the group $K$. Here $B^{ij}=(B^{-1})_{ij}$ is the inverse matrix
of the Killing form, $B_{ij}=B(X_{i},X_{j})$. It can be checked that $C_{2}$
commutes with all $X_{i}$ and therefore belongs to the center of
$\mathfrak{U}(\mathfrak{g})$. As a result $C_{2}$ acts as a multiplication by
a scalar on every irreducible representation of $\mathfrak{g}$ (the scalar
depends upon the considered representation). A basis of $\mathfrak{g}$ can be
chosen in a convenient way (see \cite{Brut Raczka}),
$B(H_{i},H_{j})=\delta_{ij}$, $B(X_{\alpha},\, X_{-\alpha})=1$, where $H_{i}$
compose a basis in $\mathfrak{h}$ whereas the vectors $X_{\alpha}$ span the
corresponding (one-dimensional) root spaces $\mathfrak{g}_{\alpha}$. All
other cross-products vanish. In consequence $X_{\alpha}$ and $X_{-\alpha}$
compose bases in, respectively, $\mathfrak{n}_{+}$ and $\mathfrak{n}_{-}$. In
this basis the second-order Casimir element reads as
$C_{2}=\sum_{\alpha>0}(X_{-\alpha}X_{\alpha}+X_{\alpha}X_{-\alpha})+\sum_{i}H_{i}^{2}$.
It is a known fact that $[X_{\alpha},\,
X_{-\alpha}]=B(X_{\alpha},X_{-\alpha})H_{\alpha}$, where $H_{\alpha}$ is an
element to which the root $\alpha\in\mathfrak{h}^{*}$ is dual under the
action of $\left.\tilde{B}\right|_{\mathfrak{h}}$. As a result, when
$\pi:\mathfrak{g}\rightarrow\mathfrak{gl}(V^{\lambda_{0}})$ is an irreducible
representation of $\mathfrak{g}$ with the highest weight $\lambda_{0}$, we
have
$\pi(C_{2})=(\lambda_{0}+2\delta,\lambda_{0})\mathbb{I}=l(\lambda_{0})\mathbb{I}$,
where
$(\cdot\,,\cdot):\mathfrak{h}^{*}\times\mathfrak{h}^{*}\rightarrow\mathbb{C}$
is the $\Ad^{*}$-invariant scalar product on $\mathfrak{h}^{*}$ defined by
the Killing form and $\delta=\frac{1}{2}\sum_{\alpha>0}\alpha$.

\section{Pure classical states and their algebraic characterization}

As stated in the Introduction by classical (pure) states we understand
a special class of generalized coherent states. The latter are defined in
terms of an irreducible representation of a group and the `origin'
- a chosen vector in the representation space. More precisely, let $\Pi:K\rightarrow End(V)$
be an irreducible, unitary representation of a Lie group $K$. We
fix a vector $v_{0}\in V$ of the unit length ($(v_{0}|v_{0})=1$)
and define the a manifold of coherent states as an orbit through $v_{0}$,
\begin{equation}
\mbox{\ensuremath{\mathcal{O}_{v_{0}}=\left\{ v(k)=\Pi(k)v_{0}|\, k\in K\right\} .}}
\end{equation}
From the quantum mechanical point of view it is more appropriate
to consider action of $K$ on the projective space $\mathbb{P}V$
rather than on $V$ itself, as in the physical interpretation of vectors
from $V$ their phase does not play a role and we use only vectors
normalized to unity. The group $K$ acts naturally on $\mathbb{P}V$
by projection of the representation $\Pi$: $\tilde{\Pi}(k)[v]=\left[\Phi(k)v\right],\,[v]\in\mathbb{P}V$
($[\cdot]:V\rightarrow\mathbb{P}V$ is a canonical projection of $V$
onto its projective space). Elements of $\left[\mathcal{O}_{v_{0}}\right]=\tilde{\Pi}(K)(v_{0})$
are called generalized coherent states (with respect to the representation
$\Pi$) of the Lie group $K$ with the origin at $v_{0}$ \cite{Perelomov}.
If $\Pi$ is an irreducible representation with the highest weight
$\lambda_{0}$ and corresponding weight vector $v_{\lambda_{0}}$,
the elements of $\tilde{\Pi}(K)(v_{\lambda_{0}})$ are called `coherent
states closest to the classical' \cite{Perelomov}. As explained above we
will call them simply (pure) classical states. They can be physically
interpreted as orthogonal projections on all pure states generated
by action of $K$ on $V^{\lambda_{0}}$. The set of pure classical
states will be denoted $CS_{K}(V^{\lambda_{0}})$.

There exists a simple, purely algebraic characterization of the set of
classical states, i.e., in the group representation language, the orbit
through the highest weight vector, given by Liechtenstein
\cite{Lichtenstein}. Let $L:V^{\lambda_{0}}\vee V^{\lambda_{0}}\rightarrow
V^{\lambda_{0}}\vee V^{\lambda_{0}}$ represents the second order Casimir
($C_{2}$) of $\mathfrak{g}$ on the symmetrization $V^{\lambda_{0}}\vee
V^{\lambda_{0}}$ of $V^{\lambda_{0}}\otimes V^{\lambda_{0}}$, achieved by
extending representation $\pi\otimes\mathbb{I}+\mathbb{I}\otimes\pi$ of
$\mathfrak{g}$ to representation of $\mathfrak{U}(\mathfrak{g})$). The result
of Lichteinstein states that
\begin{equation}
[v]\in CS_{K}(V^{\lambda_{0}})\,\Longleftrightarrow L(v\otimes v)
=(2\lambda_{0},2\lambda_{0}+\delta)v\otimes v=l(2\lambda_{0})v\otimes v\,.
\end{equation}

Since $K$ is compact all finite dimensional representations of its
complexification $G$ (and therefore also of its Lie algebra $\mathfrak{g}$)
are reducible. In particular, the symmetric tensor product decomposes into a
direct sum of irreducible representations. One of these is always the
representation $V^{2\lambda_{0}}$, and in fact this the one with the largest
value $C_{2}$, hence we can write
\begin{equation}\label{eq:ireducible components}
V^{\lambda_{0}}\vee V^{\lambda_{0}}=V^{2\lambda_{0}}
\oplus\bigoplus_{\beta<2\lambda_{0}}V^{\beta}\,,
\end{equation}
where $\beta<2\lambda_{0}$ means that the summands correspond to values of
$C_{2}$ smaller than that for $V^{2\lambda_{0}}$)

Therefore,
\begin{equation}\label{eq:Decomposition}
L=l(2\lambda_{0})\mathbb{P}_{2\lambda_{0}}\oplus
\bigoplus_{\beta<2\lambda_{0}}l(\beta)\mathbb{P}_{\beta}\,,
\end{equation}
where $\mathbb{P}_{(\cdot)}$ denote projections onto appropriate subspaces of
$V^{\lambda_{0}}\vee V^{\lambda_{0}}$. One can therefore reformulate
Lichteinstein theorem as follows,
\begin{equation}\label{eq:Lichteinstein}
[v]\in CS_{K}(V^{\lambda_{0}})\,
\Longleftrightarrow\mathbb{P}_{2\lambda_{0}}(v\otimes v)
=v\otimes v\Longleftrightarrow v\otimes v\in V^{2\lambda_{0}}.
\end{equation}

It is therefore very easy to check whether a given state is classical. All
one has to do is to verify whether $v\otimes v\in V^{2\lambda_0}$.

\section{CP maps, Jamio\l{}kowski - Choi isomorphism and Kraus decomposition}

In the following we will make use of some techniques which, to our knowledge,
were rarely employed in the theory of coherent states. They belong to linear
algebra and are of fundamental relevance in quantum information theory, or
more precisely, the theory of entanglement. Although original motivations
behind the concepts we are going to employ are of no importance for our
goals, we found it expedient to present them to some extend to make the
techniques more accessible.

A state of a quantum system interacting with an environment is described by
its density matrix $\rho$, i.e., a positive semi-definite, and thus
necessarily Hermitian, linear operator acting on the Hilbert space of the
system. In order to be able to calculate probabilities of events and quantum
averages (expectation values) of observables we impose an additional
condition of the unit trace, $\tr\rho=1$. Whatever happens to a state during
its evolution, the defining properties of positive semi-definiteness and
normalization must be retained if the evolved state has to be interpreted as
some density matrix. Thus a minimal condition which must be fulfilled by an
operator $\Lambda$ representing quantum evolution (such an operator acts in
the space of density matrices, or more generally in the space of linear
operators on the original Hilbert space) is that it transforms positive
semi-definite operators into like ones, i.e., it preserves positivity. Such
operators (the notion of map is commonly used in this context) are called
positive. A moment of reflection suffices to realize that positivity is only
a necessary but not sufficient condition requested from a map representing
quantum evolution. Indeed, the system in question can be always treated as a
part of a larger one consisting of it and some environment. The density of
state of the compound system must also evolve keeping the positive
semi-definiteness intact, even if `nothing happens' to the environment
itself. It means that tensoring $\Lambda$ with the identity (representing
lack of actual evolution of the environment) must also produce a positive map
acting on density states of the system plus environment. A map $\Lambda$
fulfilling this condition is called completely positive.

The formal definition reads,
\begin{defn*} Let $V$ be a complex finite-dimensional Hilbert space and denote
by B(V) the space of linear operators on $V$. A completely positive map on
$V$ (in short CP map on $V$) is a linear mapping $\Lambda:B(V)\rightarrow
B(V)$ between sets of linear operators on $V$ that,
\begin{itemize}
\item Preserves hermiticity,
    $A=A^{\dagger}\Rightarrow\Lambda(A)=\Lambda(A)^{\dagger}$.
\item Preserves positivity of operators,
    $\Lambda\geq0\Rightarrow\Lambda(A)\geq0$. i.e., $\Lambda$ is a
    positive map.
\item The map $\mathbb{I}_{p}\otimes\Lambda$, where $\mathbb{I}_{p}$ is
    the $p\times p$ identity matrix, is positive for arbitrary $p$.
\end{itemize}
\end{defn*}
\noindent The set of CP maps on a Hilbert space $V$ will be denoted $CP(V)$.

To make the introduced concept useful in applications we need two things, a
practical criterion allowing for an easy check whether a map is completely
positive and, possibly, a description of the structure of completely positive
maps. The former is based on an interesting connection between positive
bilinear operators on some Hilbert space $V$, i.e., positive semi-definite
linear operators on $V\otimes V$, the set of which we will denote by
$P(V\otimes V)$ and (completely) positive maps on the space of linear
operators on $V$. It is given by the so called Choi-Jamio{\l{}}kowski
isomorphism which we now briefly describe.

\begin{thm*} (Jamio\l{}kowski-Choi \cite{Zyczkowski Bengengson})
There is one to one correspondence between completely positive maps
on $N$-dimensional complex Hilbert space $V$ and positive operators
on $V\otimes V$. The isomorphism is given by the Jamio\l{}kowski mapping
$J:CP(V)\rightarrow P(V\otimes V)$,

\begin{equation}\label{eq:Jam}
J(\Lambda)=(\mathbb{I}_{N}\otimes\Lambda)(|\Phi)(\Phi|)\,,
\end{equation}
where $\Lambda\in CP(V)$, $\Phi=\sum_{i=1}^{i=N}e_{i}\otimes e_{i}$ is the
maximally entangled state in $\mathcal{H}$, and
$\left\{e_{i}\right\}_{i=1}^{i=N}$ is some fixed orthogonal basis in $V$.
\end{thm*}

Checking the complete positivity can be thus reduced to determining whether
the corresponding operator $J(\Lambda)$ acting on $V\otimes V$ is positive
semi-definite which can be easily achieved by the spectral decomposition of
$J(\Lambda)$.

In the applications that follow we will use the reasoning going in the
reverse direction. Knowing that a bilinear operator on $V\otimes V$ is
positive semi-definite we infer that the corresponding map $\Lambda$ is
completely positive. To this end we need the following
\begin{thm*}
The inverse of the Jamio{\l{}}kowki map $J^{-1}:P(V\otimes V)\rightarrow
CP(V)$ is given by
\begin{equation}\label{eq:inverse Jam}
(J^{-1}(A))(\rho)=\tr_{1}\left[(\rho^{T}\otimes\mathbb{I}_{N})\, A\right]\,,
\end{equation}
where $A\in P(V\otimes V)$, $\rho\in B(V)$, $\tr_{1}:B(V\otimes V)\rightarrow B(V)$
is the partial trace over the first Hilbert space, and $\rho^{T}$
is the transpose of the operator $\rho$ in the basis $\left\{ e_{i}\right\} _{i=1}^{i=M}$.
\end{thm*}

The second ingredient important in our argumentation is the above mentioned
structural characterization of completely positive maps. It is provided by
the fact that each CP map allows a so called Kraus decomposition
\cite{Zyczkowski Bengengson}.
\begin{thm*}
For each $\Lambda\in CP(V)$ there exists a set of operators
$T_{\alpha}:V\rightarrow V$ ($\alpha\in\mathcal{A}$, where $\mathcal{A}$ is
some set of indices) such that for all $\rho\in B(V)$,
\begin{equation}\label{eq:Kraus}
\Lambda(\rho)=\sum_{\alpha\in\mathcal{A}}T_{\alpha}\rho T_{\alpha}^{\dagger}\,.
\end{equation}
\end{thm*}
The form of $\Lambda$ given by Equation (\ref{eq:Kraus}) is called its Kraus
decomposition with Kraus operators $T_\alpha$.

The Kraus decomposition is not unique yet there is a distinguished one
associated with the spectral decomposition of $J(\Lambda)=A$. If $\left\{
f_{\alpha}\right\} _{\alpha\in\mathcal{A}}$ is the orthonormal basis of
eigenvectors of $A$ that correspond to eigenvalues $\left\{
\lambda_{\alpha}\right\} _{\alpha\in\mathcal{A}}$, we define,
\begin{equation}
T_{\alpha}=\lambda_{\alpha}^{\frac{1}{2}}(\Phi|\otimes\mathbb{I}_{N})
(\mathbb{I}_{N}\otimes|f_{\alpha})\,.
\end{equation}
The notation used in in the above formula, although commonly used, probably
needs some elucidation. Observe that both $(\Phi|$ and $|f_\alpha)$ are
linear combinations of simple tensors (the former from its definition, the
latter as an eigenvector of $A$ acting in the tensor product $V\otimes V$).
For simple tensors the corresponding formula reads
$(e_i|\otimes(e_i|\otimes\mathbb{I}_N)(\mathbb{I}_N\otimes|a)\otimes|b)=
(e_i|a)\,|b)(e_i|$, which is indeed a linear operator on $V$.

It turns out \cite{Zyczkowski Bengengson} that $T_{\alpha}$ form indeed Kraus
decomposition of $\Lambda$. The importance of this particular Kraus
decomposition is twofold. Firstly, operators $T_{\alpha}$ are orthogonal to
each other with respect to standard Hilbert-Schmidt product on $B(V)$.
Secondly, the cardinality of $\mathcal{A}$ is minimal. It is possible to
express matrix coefficients of any $A\in P(V\otimes V)$ in terms of operators
from Kraus decomposition of the CP map corresponding to it. It can be proved
\cite{Classical quantum states} that,
\begin{equation}\label{eq:Kraus expectation}
(v_{1}\otimes v_{2}|A(v_{3}\otimes v_{4}))
=\sum_{\alpha\in\mathcal{A}}(v_{1}|T_{\alpha}\mathcal{K}v_{2})
(\mathcal{K}v_{3}|T_{\alpha}^{\dagger}v_{4})\,,
\end{equation}
where $v_{1},v_{2,}v_{3},v_{4}\in V$ and $\mathcal{K}$ is the complex
conjugation of a vector expressed in the base used to define Jamio\l{}kowski
isomorphism,
$\mathcal{K}(\sum_{i=1}^{i=N}v^{i}e_{i})=\sum_{i=1}^{i=N}\overline{v^{i}}e_{i}$.

In physical applications an important class of CP maps is the class of so
called quantum channels, i.e., CP maps that preserve traces. A map $\Lambda$
is a quantum channel if $\tr\left[\Lambda(\rho)\right]=\tr\left[\rho\right]$.
On the level of Kraus decomposition of $\Lambda$ this condition reduces to
the requirement that,
$\sum_{\alpha\in\mathcal{A}}T_{\alpha}T_{\alpha}^{\dagger}=\mathbb{I}_{N}$.
How is this condition realized on the level of the operator $A=J(\Lambda)\in
P(V\otimes V)$? It is easy to check (see Eq. (\ref{eq:inverse Jam}) and Eq.
(\ref{eq:Kraus})) that it is necessary and sufficient to have
$\tr_{1}\left[A\right]=\mathbb{I}_{N}$. In this article we focus on the
situation when we have some nonnegative $A$ with only one nonzero eigenvalue.
As discussed above this situation allows to chose only one Kraus operator in
the decomposition of the corresponding $\Lambda$. If we assume that $\Lambda$
is a quantum channel we get that the corresponding Kraus operator
$T_{\alpha_{0}}$ is unitary,
\begin{equation}\label{eq:Quantum channel}
T_{\alpha_{0}}T_{\alpha_{0}}^{\dagger}=\mathbb{I}_{N}\,.
\end{equation}
Note that if $T_{\alpha_{0}}T_{\alpha_{0}}^{\dagger}\propto\mathbb{I}_{N}$
one can rescale the initial $A$ ($A\rightarrow A'=cA)$ so that resulting
$T_{\alpha_{0}}'$is unitary. By the virtue of Eq.(\ref{eq:Kraus expectation})
in the case of unitary $T_{\alpha_{0}}$ expectation value of $A$ can be
expressed in terms of antiunitary operator
$\theta=T_{\alpha_{0}}\mathcal{K}$,
\begin{equation}\label{eq:Antiunitary 1}
(v\otimes v|A(v\otimes v))=\left|(v|\theta v)\right|^{2}\,.
\end{equation}

This observation will turn out to be crucial during the discussion of
classical mixed states.

\section{Mixed classical states and their characterization}

The definition of classical states is extended to the case when state of a
system is described by a density matrix {\cite{Classical quantum
states,Classicality of spin states}}.

\begin{defn*} The set of mixed classical states on $V^{\lambda_{0}}$,
denoted in the following by $MCS_{K}(V^{\lambda_{0}})$) consists
of mixed states that can be expressed as a convex combination of projections
on pure classical states,
\begin{equation}
\rho\in MCS_{K}(V^{\lambda_{0}})\,\Longleftrightarrow\rho
=\sum_{i}p_{i}|v_{i})(v_{i}|,\,\sum_{i}p_{i}=1,
\mbox{\ensuremath{p_{i}\geq0}},\,[v_{i}]\in CS_{K}(V^{\lambda_{0}})\,.
\end{equation}
\end{defn*}
To treat pure and mixed states on the same footing we may identify a pure
state $[v]$ with the projection on $v$, i.e., $[v]\sim|v)(v|$. The set of
mixed classical states, $MCS_{K}(V^{\lambda_{0}})$, is in this language the
convex hull of the pure classical states.

In order to detect effectively classical states it would be desirable to find
an extension of the Liechtenstein criterion to to mixed states. Unfortunately
no such extension is known. One way of attacking the problem consists of
employing the so called convex roof construction. To do this we need some
results from convex geometry. Let $W$ be a real, finite dimensional vector
space. If $S$ is an arbitrary subset of $W$, by $conv(S)$ we denote convex
hull of $S$, i.e., the set composed of all convex combinations of points from
$S$, thus $conv(S)$ is the smallest convex set containing $S$. If $S$ is
compact then $conv(S)$ is also compact. Let $C$ be a convex compact subset of
$W$ and let $E$ be the set of its extremal points, i.e., points that do not
lie in the interior of any line segment contained in $C$. By the Krein-Milman
theorem $C=conv(E)$. If $f:E\rightarrow\mathbb{R}$ is a continuous function
we define its convex roof $f^{\cup}$ extension \cite{Ulhman,Grabowski},
\begin{equation}
f^{\cup}(x)=\underset{\sum_{k}p_{k}x_{k}=x}{\inf}\sum_{k}p_{k}f(x_{k})\,\,,\, x\in C,\, x_{k}\in E\,,
\end{equation}
where the infimum is taken over all possible convex decompositions of $x$
onto vectors from the set of extremal points $E$. In the following we shall
make use of the properties of $f^{\cup}$ outlined by the theorem (see
\cite{Ulhman,Grabowski}),
\begin{thm*} Function
$f^{\cup}$ is convex in $C$. Moreover, $f^{\cup}$ is the smallest
convex extension of $f$ (i.e. smallest convex function that coincides
with $f$ on $E$).
\end{thm*}
Let $E_{0}\subset E$ be some compact subset of the set of extremal points and
let $conv(E_{0})\subset C$ be its convex hull. If $f|_{E_{0}}=c$ and
$f|_{E\setminus E_{0}}>c$, then $f^{\cup}(x)=c$ if and only if $x\in
conv(E_{0})$. Therefore $f^{\cup}$ can serve as an identifier of the set
$conv(E_{0})$.

Let us come back to our main considerations. In our framework mixed states
$MS$ correspond to $C$ whereas pure states correspond to $E$. Pure classical
states $CS_{K}(V^{\lambda_{0}})$ play the role of $E_{0}$ and therefore mixed
classical states $MCS_{K}(V^{\lambda_{0}})$ can be identified with
$conv(E_{0}).$ By constructing appropriate function on the set of pure states
it is possible to construct a characterization of mixed classical states. Let
us define,
\begin{equation}
f_{1}(|\phi)(\phi|)=\sqrt{(\phi\otimes\phi|(\mathbb{I}\otimes\mathbb{I}
-\mathbb{P}_{2\lambda_{0}})|\phi\otimes\phi)}\,,
\end{equation}
where $\mathbb{P}_{2\lambda_{0}}$ is the projection onto the representation
with the highest weight $2\lambda_{0}$ embedded in $V^{\lambda_{0}}\vee
V^{\lambda_{0}}$ (see Eq.~(\ref{eq:ireducible components})). It is clear that
$f_{1}$ is well defined and continuous and reaches the minimum (equal to $0$)
on the set of pure classical states $CS_{K}(V^{\lambda_{0}})$ (see
Eq.(\ref{eq:Lichteinstein})). Therefore $f_{1}^{\cup}$ will distinguish
between classical and nonclassical mixed states. Due to the fact that $f_{1}$
is $\frac{1}{2}$ - homogenous we can write,
\begin{equation}
f_{1}^{\cup}(\rho)=\underset{\sum_{k}|v_{k})(v_{k}|
=\rho}{\inf}\sum_{k}\sqrt{(v_{k}\otimes v_{k}|(\mathbb{I}\otimes\mathbb{I}
-\mathbb{P}_{2\lambda_{0}})|v_{k}\otimes v_{k})}\,.
\end{equation}
Here the infimum is taken over all decompositions of $\rho$ into a sum of
operators of rank one (not necessary normalized). In general the infimum in
the formula for $f_{1}^{\cup}$ cannot be computed explicitly for arbitrary
$\rho$, one has then rely on various, relatively easily computable estimates,
which, however, give only sufficient criteria of non-classicality
\cite{Mintert 3} leaving a margin of uncertainty in discriminating mixed
classical states.  There are however cases when the effective computation of
the infimum is possible \cite{Ulhman}. They correspond to situations when the
operator expectation value of
$\mathbb{I}\otimes\mathbb{I}-\mathbb{P}_{2\lambda_{0}}$ can be expressed in
terms of some antiunitary operator $\tilde{\theta}$ (note that at this point
$\tilde{\theta}$ can be different from $\theta$ in Eq.(\ref{eq:Antiunitary
1})) in the following way,
\begin{equation}\label{eq:Antiunitary2}
(v\otimes v|\left(\mathbb{I}\otimes\mathbb{I}
-\mathbb{P}_{2\lambda_{0}}\right)(v\otimes v))
=\left|(v|\tilde{\theta}v)\right|^{2}\,.
\end{equation}
In such situations we have,
\begin{equation}\label{eq:single component}
f_{1}^{\cup}(\rho)=\underset{\sum_{k}|v_{k})(v_{k}|
=\rho}{\inf}\sum_{k}|(v_{k}|\tilde{\theta}v_{k})|\,,
\end{equation}
and we can perform the minimization \cite{Ulhman},
\begin{equation}\label{eq:single component 1 - exact}
f_{1}^{\cup}(\rho)=\max\left\{ 0,\mu_{1}-\sum_{j=2}^{r}\mu_{j}\right\} \,,
\end{equation}
where $\left\{ \mu_{j}\right\} _{j=1}^{j=r}$ are increasingly ordered
eigenvalues of the operator
$\left|\sqrt{\rho}\tilde{\theta}\sqrt{\rho}\right|$. According to our
knowledge situations expressed by Eq.(\ref{eq:Antiunitary2}) are the only
ones in which in is possible to compute $f_{1}^{\cup}$ explicitly. The list
of known examples of this kind described in literature \cite{Classical
quantum states,Bosons and Fermions} is short and contains only three
examples,
\begin{enumerate}
\item Three-dimensional (labeled by spin $S=1$) representation of
    $K=SU(2)$. It is a know fact that $V^{1}\vee V^{1}=V^{2}\oplus
    V^{0}$, where $V^{0}$ is the one-dimensional trivial representation
    (labeled by spin $S=0$) and $V^{2}$ is the five-dimensional
    representation (labeled by spin $S=2$). This representation is used
    used in the description of two bosons of spin $S=1$.
\item Four-dimensional representation of $SU(2)\times SU(2)$ defined by
    its natural action on
    $\mathcal{H=}\mathbb{C}^{2}\otimes\mathbb{C}^{2}$. This
    representation is used to describe entanglement of two qubits
    \cite{Classical quantum states,wootters}.
\item Six-dimensional representation of $SU(4)$ labeled by highest weight
    $(1,1,0)$ (for more details concerning notation see \cite{Alex}).
    Representation $V^{(1,1,0)}$ is isomorphic to the six-dimensional
    representation of $SU(4)$ acting on
    $\mathbb{C}^{4}\vee\mathbb{C}^{4}$. This representation is natural
    for the description of the entanglement of two fermions with spin
    $S=\frac{3}{2}$.
\end{enumerate}
It is important to note that in each of those cases there exists an
epimorphism of the appropriate group $K$ onto the group $SO(N)$,
where $N$ is the dimension of the irreducible representation of $K$.
This observation was first made in \cite{Classical quantum states}
and it was the actual inspiration for this article. In the next section
we prove that these examples are not accidental and are the manifestation
of a rather general principle relating epimorphisms of $K$ and some
$SO(N)$, antiunitary operators, and the decomposition of the symmetric
power of the representation considered onto irreducible components.

\section{Main results}

In this part we characterize in terms of the representation theory of compact
semisimple Lie groups all situations in which Eq.(\ref{eq:Antiunitary2})
takes place and computation of the {}``nonclassicality witness''
$f_{1}^{\cup}$ is possible (see Eq.(\ref{eq:single component 1 - exact})).
Let us first introduce the concept of antiunitary operators that {}``detect
classicality''. It will prove to be useful in our considerations.
\begin{defn*} We shall say that an antiunitary operator
$\theta:V^{\lambda_{0}}\rightarrow V^{\lambda_{0}}$
detects classicality if it satisfies the following,\end{defn*}
\begin{itemize}
\item $\theta$ is $K$-invariant, that is,
    $\Pi(k)^{\dagger}\theta\Pi(k)=\theta$ for each $k\in K$.
\item Expectation value of $\theta$ vanishes exactly on classical states,
    $(v|\theta v)=0\Longleftrightarrow[v]\in CS_{K}(V^{\lambda_{0}})$.
\end{itemize}
We present our results in two theorems. First of them relate the existence of
antiunitary operators detecting classicality to the decomposition of
$V^{\lambda_{0}}\vee V^{\lambda_{0}}$. Second theorem connects this kind of
antiunitary operators with the existence of epimorphisms of the group $K$
onto some orthogonal group.

\begin{thm*}\label{Theorem1} Let $K$ be the semisimple,
compact and connected Lie group. Let $\Pi$ be some irreducible unitary
representation of the group $K$ in the Hilbert space $V^{\lambda_{0}}$ with
the highest weight $\lambda_{0}$. The following two statements are
equivalent,\end{thm*}
\begin{description}
\item{1.\ } There exists an antiunitary operator
    $\theta:V^{\lambda_{0}}\rightarrow V^{\lambda_{0}}$ detecting
    classicality.
\item{2.\ } $V^{\lambda_{0}}\vee V^{\lambda_{0}}=V^{2\lambda_{0}}\oplus
    V^{0}$, where $V^{0}$ is a trivial representation of the group $K$.
\end{description}
\begin{proof} ($1\rightarrow2$) Let $\theta=T\tilde{K}$ where $T$
is an unitary operator and $\tilde{\mathcal{K}}$ is the operator of the
complex conjugation is some fixed basis of $V^{\lambda_{0}}$, say $\left\{
e_{i}\right\} _{i=1}^{i=N}$. Define an operator $A\in
P(V^{\lambda_{0}}\otimes V^{\lambda_{0}})$ as an image of the Jamio\l{}kowski
map (defined with respect to the basis $\left\{ e_{i}\right\} _{i=1}^{i=N}$)
of the CP map $\Lambda(\rho)=T\rho T^{\dagger}$ (see Eq.(\ref{eq:Jam})). It
is easy to check that matrix elements of $A$ are given by the following
formula (see Eq.(\ref{eq:Kraus expectation}) and \cite{Kotowscy})
\begin{equation}
(v_{1}\otimes v_{2}|A(v_{3}\otimes v_{4}))
=(v_{1}|T\tilde{\mathcal{K}}v_{2})(T\tilde{\mathcal{K}}v_{3}|v_{4})
=(v_{1}|\theta v_{2})(\theta v_{3}|v_{4})\,.
\end{equation}
We now claim that the operator $A$ is proportional to
$\mathbb{I}\otimes\mathbb{I}-\mathbb{P}_{2\lambda_{0}}$ (when both operators
are understood as acting on $V^{\lambda_{0}}\vee V^{\lambda_{0}}$). Indeed,
$A$ is symmetric and nonnegative. It is also $K$-invariant due to the
$K$-invariance of $\theta$,
\begin{eqnarray}\label{eq:Long}
(\Pi(k)v_{1}\otimes\Pi(k)v_{2}|A(\Pi(k)v_{3}\otimes\Pi(k)v_{4}))
=(\Pi(k)v_{1}|\theta\Pi(k)v_{2})(\theta\Pi(k)v_{3}|\Pi(k)v_{4})\nonumber \\
=(v_{1}|\theta v_{2})(\theta v_{3}|v_{4})
=(v_{1}\otimes v_{2}|A(v_{3}\otimes v_{4}))\,.
\end{eqnarray}
Thus,
\begin{equation}
A=a_{2\lambda_{0}}\mathbb{P}_{2\lambda_{0}}
\oplus\bigoplus_{\beta<2\lambda_{0}}a_{\beta}\mathbb{P}_{\beta}\,,
\end{equation}
where the above sum corresponds to the decomposition of $V^{\lambda_{0}}\vee
V^{\lambda_{0}}$ onto irreducible components (see Eq.(\ref{eq:ireducible
components})) and $a$'s are some nonnegative scalars. By definition $A$ has
only one eigenvector (see our remarks during the discussion of the CP maps
and the Kraus decomposition). Projection on this eigenvector cannot belong to
$V^{2\lambda_{0}}$ because the expectation value of $\theta$ vanish on
coherent states. On the ofter hand, by the theorem of Lichtenstein and the
properties of $\theta$,
\begin{equation}
|(v|\theta v)|>0\Longleftrightarrow
\bigoplus_{\beta<2\lambda_{0}}\mathbb{P}_{\beta}(v\otimes v)\neq0\,.
\end{equation}
From this it follows that there is only one $\beta$ in the above sum and the
corresponding $\mathbb{P}_{\beta}$ is one dimensional because $A$ has rank
one. Therefore we get a one dimensional irreducible representation of group
$K$. But this irreducible representation must be a trivial representation due
to the fact that $K$ is semisimple.

($2\rightarrow1$) If $V^{\lambda_{0}}\vee V^{\lambda_{0}}=V^{2\lambda_{0}}\oplus V^{0}$
operator $\mathbb{I}\otimes\mathbb{I}-\mathbb{P}_{2\lambda_{0}}=\mathbb{P}_{0}$
(acting on $V^{\lambda_{0}}\vee V^{\lambda_{0}}$) has rank one and
is nonnegative. If we apply to it the inverse of the Jamio\l{}kowski
isomorphism (with respect to some fixed basis $\left\{ e_{i}\right\} _{i=1}^{i=N}$)
we get $\Lambda(\mathbb{P}_{0}$) together with the corresponding
Kraus operator $T$ (see Eq.(\ref{eq:inverse Jam}) and Eq.(\ref{eq:Kraus})).
By the Eq.(\ref{eq:Kraus expectation}) we have,

\begin{equation}\label{eq:P_0}
(v_{1}\otimes v_{2}|\mathbb{P}_{0}(v_{3}\otimes v_{4}))=(v_{1}|T\mathcal{K}v_{2})(T\mathcal{K}v_{3}|v_{4})\,,
\end{equation}
 where $\mathcal{K}$ is the complex conjugation in the basis $\left\{ e_{i}\right\} _{i=1}^{i=N}$.
We claim that the antilinear operator $\theta=T\mathcal{K}$ is proportional
to the antiunitary operator detecting classicality. By the equation
(\ref{eq:P_0}) $\theta$ is $K$-invariant. It follows from the
$K$-invariance of $\mathbb{P}_{0}$ and the calculation is essentially
the reverse of the calculation in Eq.(\ref{eq:Long}). Because of
the decomposition $V^{\lambda_{0}}\vee V^{\lambda_{0}}=V^{2\lambda_{0}}\oplus V^{0}$
and Eq.(\ref{eq:P_0}) we have $(v|\theta v)=0\Longleftrightarrow[v]\in CS_{K}(V^{\lambda_{0}})$.
The only thing that needs be proved is that $T$ can be rescaled to the
unitary operator. This follows from the discussion of the relation
between nonnegative operators on the product of Hilbert spaces and
quantum channels (see Eq.(\ref{eq:Quantum channel})). The necessary
and sufficient condition for $T$ to be proportional to the unitary
operator is $\tr_{1}\left[\mathbb{P}_{0}\right]\propto\mathbb{I}_{N}$.
$\mathbb{P}_{0}$ is the orthogonal projection onto one dimensional
trivial representation $V^{0}$ in the decomposition of $V^{\lambda_{0}}\vee V^{\lambda_{0}}$.
It can be thus written in the form of the integral with respect to
the Harr measure $\mu$ over the whole $K$ \cite{Brut Raczka},
\begin{equation}
\mathbb{P}_{0}=\int_{K}\Pi(k)\otimes\Pi(k)d\mu(k)\,.
\end{equation}
As a result we have,
\begin{equation}
\tr_{1}\left[\mathbb{P}_{0}\right]=\int_{K}\tr\left[\Pi(k)\right]\Pi(k)d\mu(k)=\int_{K}\chi_{\lambda_{0}}(k)\Pi(k)d\mu(k)\,,
\end{equation}
 where $\chi_{\lambda_{0}}(k)$ is the character of the representation
$\Pi$. By the general representation theory of compact Lie groups
we have (see \cite{Brut Raczka}),
\begin{equation}
\int_{K}\chi_{\lambda_{0}}(k)\Pi(k)d\mu(k)=\frac{\mathbb{I}_{N}}{N}\,.
\end{equation}
Therefore the proof in now completed. \end{proof}

Note that in the assumptions of the above theorem there is no reference to
the dimension of the considered representation $V^{\lambda_{0}}$. It is
nevertheless clear that when $\dim(V^{\lambda_{0}})=1$ and
$\dim(V^{\lambda_{0}})=2$ both statements that are meant to be equivalent are
at the same time false.

The above proved theorem states that cases when operator
$\mathbb{I}\otimes\mathbb{I}-\mathbb{P}_{2\lambda_{0}}$ has rank one
correspond exactly to the appearance of antiunitaries that detect
classicality. As advertised, it turns out that such cases are related to the
existence of an epimorphism between the group $K$ and one of three groups: $SO(N)$ (
for $N=\dim(V^{\lambda_{0}})$), $G_2$ or $Spin(7)$. We formulate this fact in the following
theorem.

\begin{thm*}\label{Theorem2} Let $K$ be a semisimple,
compact and connected Lie group. 
The following two statements are equivalent,\end{thm*}
\begin{description}
\item{1.\ } There exists an irreducible unitary
representation $\Pi$ of the group $K$ in the Hilbert space $V^{\lambda_{0}}$ 
with the highest weight $\lambda_{0}$ ($N=\dim(V^{\lambda_{0}})>2$). On   $V^{\lambda_{0}}$ there exists an antiunitary operator
    $\theta:V^{\lambda_{0}}\rightarrow V^{\lambda_{0}}$ detecting
    classicality. 
\item{2.\ } There exists an epimorphism $h:K\rightarrow SO(N)$, or $h:K\rightarrow G_2$ (the exceptional
    Lie group $G_2$ \cite{Adams}) with $N=7$, or $h:K\rightarrow Spin(7)$
    with $N=8$.
\end{description}
\begin{proof} ($1\rightarrow2$) Because $\theta$ is antiunitary
it is possible to chose the orthonormal basis $\left\{ e_{i}\right\}
_{i=1}^{i=N}$ of $V^{\lambda_{0}}$ is such a way that each vector from the
basis is an eigenvector of $\theta$ with an eigenvalue $1$: $\theta
e_{i}=e_{i}\,,\, i=1,\ldots,N$. In this basis $\theta$ act as a complex
conjugation,
\begin{equation}
v=\sum_{i=1}^{i=N}v_{i}e_{i}\Longrightarrow\theta v
=\sum_{i=1}^{i=N}\bar{v}_{i}e_{i}\,.
\end{equation}
From the $K$-invariance of $\theta$ it follows that in this basis unitary
operators defining $\Pi(k)$ are real and therefore also orthogonal. Because $\Pi$ is
continuous the image of a connected group $K$ must be connected and therefore
$\Pi$ defines a homomorphism $h:K\rightarrow SO(N)$. Note that each
representative $v\in V^{\lambda_{0}}$ of a state
$[v]\in\mathbb{P}V^{\lambda_{0}}$ can be decomposed into its real and
imaginary part, $v=u+i\cdot w$ in the basis $\left\{ e_{i}\right\}
_{i=1}^{i=N}$ described above, i.e., with the vectors $u$ and $w$ being real
linear combinations of basis vectors. For a classical state in such a form we
have $0=(v|\theta v)=(u|u)-(w|w)+2i(u|w)$, hence all classical states are
represented by vectors $v=u+i\cdot w$, where $(u|u)=(w|w)=1$ and $(u|w)=0$.
In particular, $v_{\lambda_{0}}=u_{0}+i\cdot w_{0}$ for some orthogonal and
appropriately normalized $u_{0}$ and $w_{0}$. For $\tilde{x}\in
CS_{K}(v_{\lambda_{0}})$ let$\tilde{v}=\tilde{u}+i\cdot\tilde{w}$ be its
representative normalized as above. We claim that $\tilde{u}=\Pi(k)u_{0}$ and
$\tilde{w}=\Pi(k)w_{0}$ for some $k\in K$. Indeed, we must have
$\tilde{v}=\epsilon\Pi(k_{1})v_{\lambda_{0}}$ for some unimodular $\epsilon$
and $k_{1}\in K$. Because representation $\Pi$ is non-trivial we can chose
$k_{2}\in K$ such that $\epsilon v_{\lambda_{0}}=\Pi(k_{2})v_{\lambda_{0}}$.
Therefore, if we take $k=k_{1}k_{2}$, we have the desired result (matrices
corresponding to $\Pi(k)$ are real in the considered basis),
\begin{equation}
\Pi(k)u_{0}+i\cdot\Pi(k)w_{0}=\Pi(k)v_{\lambda_{0}}
=\epsilon\Pi(k_{1})v_{\lambda_{0}}=\tilde{v}=\tilde{u}+i\cdot\tilde{w\,.}
\end{equation}

Thus, it is possible to generate all pairs of orthonormal vectors by the
action of $h(K)$ on vectors $u_{o}$ and $w_{0}$, i.e.\ $\Pi(K)$ acts
transitively on pairs of orthonormal vectors. This fact suffices to prove
that $h(K)$ equals $SO(N)$, $G_2$ or $Spin(7)$. In order to see this we refer
to the classical result of Montgomery and Samelson \cite{Montgomery} that
classifies all compact and connected Lie groups acting transitively and
effectively on $M$ dimensional spheres. In our case $M=N-1$ and we are
interested only in compact and connected matrix groups ($h(K)$ is obviously
compact, connected and acts transitively and effectively on the sphere). The
list of such groups is short and consists of seven cases: $SO(N)$ itself, its
three proper subgroups: $SU(\frac{N}{2}),\, Sp(\frac{N}{4})$, and
$Sp(1)\times Sp(\frac{N}{4})$ (where $Sp(\cdot)$ denotes a compact symplectic
group), $G_2\subset SO(7)$, $Spin(7)\subset SO(8)$, and $Spin(9)\subset SO(16)$. We first consider last three ``exceptional'' cases. Groups $G_2$, $Spin(7)$ and $Spin(9)$ act transitively on respectively $6$, $7$ and $15$ dimensional spheres. Those actions come from the following (faithful) representations: defining representation of $G_2$, $8$ dimensional spinor representation of $Spin(7)$ and $16$ spinor representation of $Spin(9)$. Actions of $G_2$ and $Spin(7)$ are transitive on orthonormal pairs of vectors (see \cite{Adams}, p.\ 32). Therefore those groups are permissible. On the other hand, it is known \cite{klyachko} that the $16$ dimensional representation of $Spin(9)$ does not have the desired property. Let us now consider the special unitary and symplectic subgroups of $SO(N)$. Those groups can appear only when $2$ (in the case of $SU(\frac{N}{2})$) or
$4$ (in the case of $Sp(\frac{N}{4})$ and $Sp(1)\times Sp(\frac{N}{4})$) are
divisors of $N$. Therefore when $N$ is odd the proof is finished. Now assume
that $2$ or $4$ divide $N$. Since $h(K)$ acts transitively on orthonormal
pairs of vectors, a subgroup $stab(u_{0})$ of $h(K)$ that stabilizes the
vector $u_0$ must act transitively on $S^{N-2}\cong S^{N-1}\cap
u_{0}^{\perp}$, where $u_{0}^{\perp}$ is the orthogonal complement of
$u_{0}$. We can now apply the theorem of Montgomery and Samelson for the
dimension $N-1$. Since $N-1$ is now odd we infer that $stab(u_0)=SO(N-1)$. As
a consequence we have
$dim(h(K))\geq\frac{\left(N-1\right)\left(N-2\right)}{2}=\dim(SO(N-1))$.
Since the dimensions of $SU(\frac{N}{2})$, $Sp(\frac{N}{4})$ and $Sp(1)\times
Sp(\frac{N}{4})$ are, respectively, $\frac{N^{2}}{4}-1$,
$\frac{N}{4}(\frac{N}{2}+1)$ and $\frac{N}{4}(\frac{N}{2}+1)+3$, we can exclude those groups. At the and we conclude that only
possibilities are that $h(K)=SO(N)$, $h(K)=G_2$ (when $N=7$) or $h(K)=Spin(7)$ (when $N=8$).

($2\rightarrow1$) We treat groups $SO(N),$ $G_{2}$ and $Spin(7)$ together. We consider defining representations of $SO(N)$ and  $G_{2}$ and the 8 dimensional spinor represenation of $Spin(7)$.  We
shall show that symmetric powers of those irreducible faithful representations (clearly those are also irreducible representations of the group $K$) decompose onto two ingredients: $V^{\lambda_{0}}\vee
V^{\lambda_{0}}=V^{2\lambda_{0}}\oplus V^{0}$. Then, combining this with the theorem
$6.1$ we conclude the existence of the antiunitary operator $\theta$ that
detects classicality for each of considered representations. To prove the above
decomposition we notice that each representation respects the Euclidean structure in
the relevant $V^{\lambda_{0}}$ (when viewed as subgroups of $SO(N)$, $SO(7)$
and $SO(8)$ accordingly) so $V^{\lambda_{0}}\simeq(V^{\lambda_{0}})^{*}$. As
a result we have $V^{\lambda_{0}}\vee V^{\lambda_{0}}\simeq
V^{\lambda_{0}}\vee(V^{\lambda_{0}})^{*}\simeq SEnd(V)$
($SEnd(V^{\lambda_{0}})$ denotes symmetric endomorphisms of
$V^{\lambda_{0}}$). There is one distinguished element of
$SEnd(V^{\lambda_{0}})$ that is preserved by the group action: $\mathbb{I}$ - the 
identity in $V^{\lambda_{0}}$. This element corresponds to the one
dimensional invariant subspace subspace $V^{0}$. $V^{2\lambda_{0}}$ is
realized by the the traceless operators from $SEnd(V)$. It is easy to check
that this subspace is an irreducible representation - it follows from the
transitivity of the action of the each group on pairs of orthonormal
vectors (when treated as a subgroup of $SO(N)$, $SO(7)$ and $SO(8)$
respectively). We have thus proved the decomposition $V^{\lambda_{0}}\vee
V^{\lambda_{0}}=V^{2\lambda_{0}}\oplus V^{0}$ which finishes the proof.\end{proof}

\section{Concluding remarks}
We presented group theoretical conditions for the cases when antiunitary
operator detecting classicality exists and it is possible to compute
$f_{1}^{\cup}$ exactly. Theorem~\ref{Theorem1} links such cases with
situations in which symmetric power of the considered representation
decomposes onto two irreducible components one of which is a trivial, one
dimensional representation.

In the course of the proof of the Theorem~\ref{Theorem2} we referred to the
classical work by Montgomery and Samelson \cite{Montgomery}. Although it may
seem to be a trick from a rather `high floor' we would like to stress that
the problem is not as easy as it may seem at the first sight. It turns out
that when $N$ is even there are proper subgroups of $SO(N)$ that act
transitively on $S^{N-1}$ (it turns out that this fact is directly related to
the classification of the holonomy groups of irreducible non locally
symmetric Riemannian spaces \cite{Olmos}). Nevertheless our assumption about
the existence of an `antiunitary operator detecting classicality' is strong
enough to guarantee that the image of the homomorphism we consider is the
whole $SO(N)$, $G_2$ or $Spin(7)$.

The groups enumerated in Theorem~\ref{Theorem2} appear in the context of
entanglement in the paper of Klyachko \cite{klyachko}. This is not entirely
accidental. `Systems in which all unstable states are coherent' considered by
Klyachko in his paper can be, in fact, equivalently characterized by our
Theorem~\ref{Theorem1}.

\section*{Acknowledgments}

We gratefully acknowledge supports from the Polish Ministry of Science and
Higher Education through the project no.\ N N202 090239 and the Deutsche
Forschungsgemeischaft through the grant SFB-TR12. We thank Jan Gutt,
Miko{\l}aj Rotkiewicz, and Adam Sawicki for very fruitful hints and comments.

\end{document}